\documentclass[aps,prb,preprint,showpacs,groupedaddress]{revtex4-1}  % for double-spaced preprint
\usepackage{graphicx}  % needed for figures
\usepackage{dcolumn}   % needed for some tables
\usepackage{bm}        % for math
\usepackage{amssymb}   % for math
\begin{document}

% The following information is for internal review, please remove them for submission
\leftline{Version 4 as of \today} 
\leftline{Primary author: Simon Hale}

% the following line is for submission, including submission to the arXiv!!
%\hspace{5.2in} \mbox{Fermilab-Pub-04/xxx-E}

\title{Many-body effects on the capacitance of multilayers made from strongly correlated materials}
% LIST_OF_AUTHORS_R2.TEX                 10/14/09           
%
\author{S. T. F.~Hale$^{1}$}
\author{J. K. Freericks$^{1}$}
\affiliation{$^{1}$Department of Physics, Georgetown University, Washington, D.C 20057, USA}
  % input author list
\date{\today}

\begin{abstract}
Recent work by Kopp and Mannhart\cite{KoMa} on novel electronic systems formed at oxide interfaces has shown interesting effects on the capacitances of these devices. We employ inhomogeneous dynamical mean-field theory to calculate the capacitance of multilayered nanostructures.  These multilayered nanostructures are composed of semi-infinite metallic leads coupled via a strongly correlated dielectric barrier region. The barrier region can be adjusted from a metallic regime to a Mott insulator through adjusting the interaction strength. We examine the effects of varying the barrier width, temperature, potential difference, screening length, and chemical potential. We find that the interaction strength has a relatively strong effect on the capacitance, while the potential and temperature show weaker dependence. 

\end{abstract}

\maketitle 

%\section{\label{sec:level1}First-level heading}
% sections are not used for PRL papers

\begin{center}
\textbf{I. INTRODUCTION}
\end{center}

	As the capabilities of electronic components increase and their sizes decrease, new ideas are needed to continue the advancements in technology{\cite{MeChDa}}. One of the fundamental electronic components is the capacitor, in its basic form it is a dielectric ($\kappa$) layer separating two conducting layers, which builds and stores charge on each conducting plane when an external potential is applied. Driving the development of high-performance capacitors is the further miniaturization of various electronic devices including metal-oxide semiconductor field-effect transistors (MOSFETs). In order to continue the trend of development, high-$\kappa$ insulating materials are utilized in MOSFETs and capacitors{\cite{ScGuDa,WiWaAn,Ro}}. As the size of these devices decrease, quantum mechanical effects play a greater role and complicate the trend of using higher and higher $\kappa$ materials. Recent theory work by  Kopp and Mannhart{\cite{KoMa}} and subsequent experimental work by Li \textit{et al} {\cite{Li}} has introduced the idea of using ultrathin strongly correlated electronic materials to produce controllable small or large capacitances instead of the traditional high-$\kappa$ dielectric approach. This work builds on the growing number of possible applications of oxide interfaces {\cite{ohtomo, thiel, Breit,MaSc}}, as oxides forming two dimensional electron gases at the interface present a strong candidate for capacitance enhancements. The experimental work {\cite{Li}} found a greater than 40\% increase in the gate capacitance when the mobile electrons were nearly depleted in a LaAlO$_{3}$/SiTiO$_{3}$ interface. This increase is attributed to a negative compressibility of the interface electron system.  Motivated by this work, we want to theoretically investigate the strong correlation effects on capacitance when the barier is a Mott insulator. 

	 We focus on constructing theoretical nanostructure devices consisting of ballistic metal leads on both sides of a strongly correlated electron dielectric layer. Inhomogeneous dynamical mean-field theory{\cite{pottnolt}} (IDMFT) the theoretical framework for this work, allows for the self-consistent calculation of the properties of such devices. We use the Falicov-Kimball model{\cite{Falkim}} to govern the interaction and use a Potthoff-Nolting{\cite{pottnolt}} technique for solving the IDMFT. We work in the static limit with no current flow, where all calculations can be carried out in equilibrium{\cite{Lutt}}.

The capacitance is calculated for various parameters, showing the strongest dependence on the interaction strength and weaker dependence on the temperature and applied potential. Two methods for calculating the capacitance are discussed in the paper, one based on the center of charge approach of Lang and Kohn{\cite{KoLa}} (where one measures the total charge on the capacitor) and the other on the voltage profiles through the capacitor due to Mead{\cite{Mead}} (where one measures the voltage difference between the plates). 

The general organization of the rest of this paper is as follows; in Section II, we detail the mathematical formalism and numerical issues associated with the calculation of the capacitance from the IDMFT approach. In Section III, we present numerical results of the capacitance dependence on various parameters including temperature, thickness, and dielectric screening length. We summarize the work, discussing the results and future ideas in Section IV.

\begin{center}
\textbf{II. FORMALISM}
\end{center}

	The general equation for the capacitance {\cite{jackson}}, C, of two electrodes possessing charges of $Q$ and $-Q$, separated by voltage $V$ is
\begin{equation}
C=\frac{Q}{V}.
\label{eq:cqv}
\end{equation}
	In this work, we are concerned with parallel plate capacitors; an arrangement of two-dimensional layers stacked in a sandwich configuration. Classically, the capacitance of two-plate capacitor is defined as 
\begin{equation}
C=\frac{\epsilon_{0} \epsilon_{r}A}{d},
\label{eq:geom}
\end{equation}
where $\epsilon_{r}$ is the relative dielectric constant of the material separating the plates, $\epsilon_{0}$ ($\kappa=\epsilon_{r}$) is the dielectric constant of the vacuum, $A$ is the area of the plates, and $d$ is the thickness of the dielectric. These equations assume that the charges sit on an idealized surface plane of zero thickness. This assumption does not hold in reality and the electron density distribution must be taken into account. Kohn and Lang{\cite{KoLa}} showed that the effective position of the lead surface, $z_{0}^{R,L}$, can be calculated from a center of charge approach, 
\begin{equation}
z_{0}^{L}=\sum_{\alpha=-\infty}^{center} z_{\alpha} \rho_{\alpha} / \sum_{\alpha=-\infty}^{center}\rho_{\alpha},  %\int _{-\infty}^{0} z \rho (z;0) dz/ \int _{-\infty}^{0}\rho (z;0) dz,
\end{equation}
where $z_{\alpha}$ is the position of plane $\alpha$ in the $z$-direction, and $\rho_{\alpha}$ is the charge density distribution on plane $\alpha$ (with $\alpha=-\infty$ being the left most plane and $\alpha = center$ being the center of the barrier region). The sum ranges over one half of the capacitor only and $z_{0}^{R}$ is equivalently defined with sum from the center of the barrier to last plane on the right.
 Equation (\ref{eq:geom}) is therefore modified and the capacitance per unit area of a two plate capacitor becomes 
\begin{equation}
\frac{C_{CoC}}{A}=\frac{\epsilon_{0} \epsilon_{r}}{(z_{0}^{R}-z_{0}^{L})}.
\label{eq:geom2}
\end{equation}
 For the rest of the paper we will refer to Eq. (\ref{eq:geom2}) as the center of charge (CoC) capacitance.

\begin{figure}[htp]
	\centering
		\includegraphics[width=120mm]{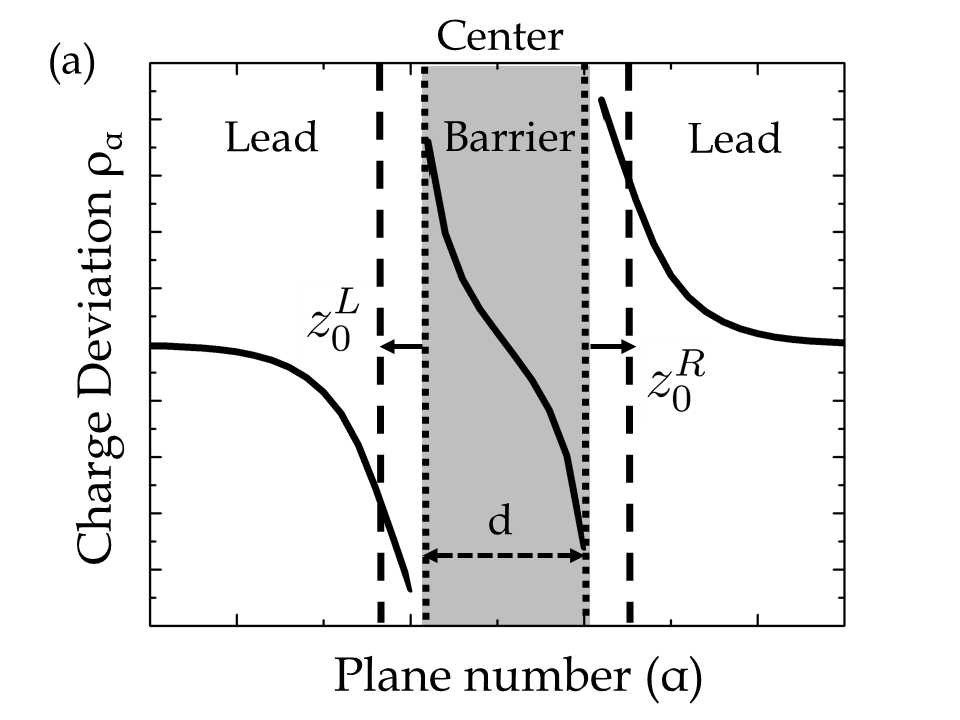}
		\includegraphics[width=120mm]{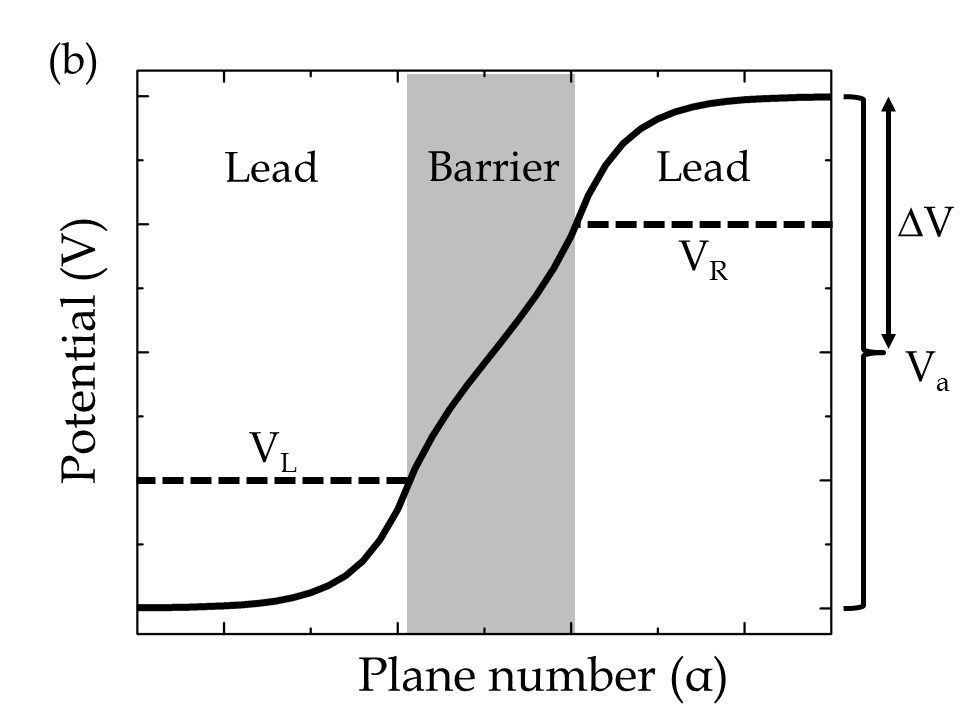}
	\caption{Schematics of the two methods of calculation of the capacitance. Panel (a) shows the center of charge (CoC) method. The parameter $z_{0}^{L}$ is calculated by using the charge distribution to the left of center and is shown as a dashed line. This line is shifted away from the original interface location shown as a dotted line. Similarly $z_{0}^{R}$ is calculated by summing the charge distribution to the right of center.  Figure (b) shows the voltage profile method (VP). $V_{L}$ is the value of the potential at the left interface and $V_{R}$ the potential at the right interface. $V_{a}$ is the applied voltage which equals two times $\Delta V$. }
\label{fig:CapVPCoC}
\end{figure}

	In addition to the assumption of an ideal surface charge, as the size of these devices approaches an ultrasmall regime Eq. (\ref{eq:geom}) will also begin to break down.  For example, a device where the effects of a single electron plays a dominate role, the calculations need to include quantum mechanics and a classical approach will not give a complete description. Quantum-mechanical effects will play an important role even before the single electron limit is reached. If the device is thin enough that the electric field is screened over a significant portion of the barrier region, there can be a noticeable reduction in the capacitance. This effect was first observed by Mead{\cite{Mead}} for thin films and gives us a modified equation{\cite{KuUl}} for a thin film parallel plate capacitance per unit area; 
\begin{equation}
\frac{A}{C_{VP}}=(d/\epsilon_{0} \epsilon_{r})\left\{1+\frac{V_{a}-(V_{R}-V_{L})}{(V_{R}-V_{L})}\right\},
\label{eq:modcap}
\end{equation}
where $V_{a}$ is the applied potential, $V_{L}$ and $V_{R}$ are the potentials at the left and right interfaces, respectively, with $(V_{R}-V_{L})$ being the potential difference across the dielectric. Written in this form the modification to the geometric capacitance can be seen as the second term in the braces. We refer to  Eq. (\ref{eq:modcap}) as the voltage profile (VP) capacitance for the remainder of this paper. A schematic representation of the two methods is shown in Fig. {\ref{fig:CapVPCoC}}.

The $C_{VP}$ method can be thought of as fixing the distance between the plates to the physical difference of the plate geometry and calculating the effective potentials at that separation. While the $C_{CoC}$ method sums the total charge, $Q$, for the two halves and calculates an effective distance between charges. Hence, the effective charge in the VP method will not equal the total charge in the CoC method, while the distance between the plates in the CoC method will not equal the distance between the plates in the VP method unless the two results coincide.  Each method has its relative strengths and relation to experiments discussed later in this paper.

		Extracting physical properties from inhomogeneous multilayered nanostructures is made possible by utilizing the algorithm originally employed by Potthoff and Nolting {\cite{pottnolt}} and later adapted by Freericks{\cite{freejk}}. This so called quantum zipper algorithm reduces the complexity of the system by stacking two-dimensional translationally  invariant planes, thereby building the inhomogeneity in the longitudinal third dimension ($z$) only. The $z$-coordinate remains in real space, while the $x$ and $y$ coordinates are Fourier transformed to wavevectors $k_{x}$ and $k_{y}$, respectively, forming a mixed basis. Greek letters ($\alpha, \beta, \gamma,...$) are used to denote the planar index of the $x-y$ planes stacked in the $z$-direction. We are left with a quasi-one-dimensional problem for each two-dimensional band energy, that can represented tridiagonally in real space and solved via the quantum zipper algorithm. The many-body equations are iterated to achieve a self-consistent solution. This Potthoff-Nolting approach is used to extract the electronic charge on each two-dimensional plane via the Green's functions {\cite{econ}}. The charges are then used in a classical calculation to find the potentials on each plane and then the local electrochemical potential. The approach is iterated until it reaches a steady state charge distribution. 
		
	We begin the calculations by defining the governing Hamiltonian for our system. The Hamiltonian involves a hopping term for the electrons and an interaction term for the sites within the barrier region. For the interaction in our numerical calculations, we employ the Falicov-Kimball model {\cite{Falkim}} which involves an interaction between spinless conduction electrons and spinless localized electrons. When the conduction electron hops onto a site already occupied by the localized electron, both electrons feel a mutual repulsion. When this correlation strength is large enough in the Falicov-Kimball model it has a Mott-like metal-insulator transition. Additionally the Falicov-Kimball model has the advantage of simplifying the solution of the IDMFT equations for this system. It turns out that the stabilization of the voltage profile under the iterative solution of the IDMFT equations is difficult, and to date has only been able to be achieved in numerical solutions that are extremely accurate, like the Falicov-Kimball model. In the second quantization formalism, the spinless Falicov-Kimball Hamiltonian{\cite{Falkim}} is, 
\begin{eqnarray}
H=-\sum_{\alpha}\sum_{i,j\in plane}t_{\alpha ij}c^{\dagger}_{\alpha i}c_{\alpha j}
-\sum_{\alpha}\sum_{i\in plane}t_{\alpha \alpha+1}\left(c^{\dagger}_{\alpha i}c_{\alpha+1 i}+c^{\dagger}_{\alpha+1 i}c_{\alpha i}\right)\nonumber\\
-\sum_{\alpha}\sum_{i\in plane} (\mu_{\alpha}-V_{\alpha}+\Delta V_{\alpha})c^{\dagger}_{\alpha i}c_{\alpha i}+
\sum_{\alpha}\sum_{i\in plane} U_{\alpha}c^{\dagger}_{\alpha i}c_{\alpha i}\left(w_{\alpha i}-\frac{1}{2}\right),
\end{eqnarray}
where the first two terms of the Hamiltonian describe intraplane and interplane electron hopping, respectively, where $c^{\dagger}_{\alpha i}$ and $c_{\alpha i}$ are creation and annihilation operators on plane $\alpha$ and site $i$, respectively. The third term describes the charge reconstruction that occurs due to an externally applied potential, $\mu_{\alpha}$ is the chemical potential on plane $\alpha$, $V_{\alpha}$ is the potential energy on plane $\alpha$ due to the Coulomb interaction of the electronic charge reconstruction, and $\Delta V_{\alpha}$ is the input applied potential on plane $\alpha$. The last term is the interaction term where $U_{\alpha}$ represents the interaction strength on plane $\alpha$, and $w_{\alpha i}$ is a classical variable that equals one if there is a localized particle at site $i$ on plane $\alpha$ and zero if there is no localized particle at site i  on plane $\alpha$.    
 
 We will use Green's functions to solve the problem. The equilibrium Green's function, in real space, and imaginary time is defined by
\begin{equation}
G_{\alpha\beta ij}(\tau)=-\left\langle T_{\tau} c_{\alpha i}(\tau) c^{\dagger}_{\beta j} (0) \right\rangle,
\label{eq:greens}
\end{equation}
for imaginary time $\tau$, where $T_{\tau}$ represents the time ordering operator. The notation $\left\langle X \right\rangle$ denotes the trace, ${\rm Tr }\  \exp(-\beta H) X$ divided by the partition function $\cal{Z}$$ ={\rm Tr }\   \exp(-\beta H )$, and the operators are expressed in the Heisenberg representation $X(\tau)=\exp(\tau H) \  X \  \exp(-\tau H)$, all with respect to the Hamiltonian $H$.
To properly express the Green's functions for the Matsubara frequencies we use a Fourier transformation
\begin{equation}
G_{\alpha\beta}(i\omega_{n})=T \int^{\beta}_{0}d\tau  e^{i\omega_{n}\tau} G_{\alpha\beta}(\tau).
\label{eq:GInt}
\end{equation}
where $T=1/ \beta$ is the temperature.

	To build our model, we need to solve for the local Green's function on each plane, which we do by employing the quantum zipper algorithm{\cite{freejk}}, based on the Potthoff-Nolting formalism{\cite{pottnolt}}. When solving for the local Green's functions, we use the fermionic Matsubara frequencies, $i \omega_{n}=i \pi T (2n+1)$. We start with the unperturbed equilibrium equation of motion (EOM),
\begin{eqnarray}
\sum_{\gamma}G_{\gamma\beta}(i \omega_{n};\bm{k}^{||})\left[ \left( i \omega_{n}+\mu_{\alpha}-V_{\alpha}+\Delta V_{\alpha}-\epsilon^{\alpha}_{\bm{k}^{||}}\right) \delta_{\alpha\gamma}+\left(t_{\alpha-1\alpha}\delta_{\gamma\alpha-1}+t_{\alpha+1\alpha}\delta_{\gamma\alpha+1}\right)
-\Sigma_{\alpha}(i \omega_{n})\delta_{\alpha\gamma}\right]=\delta_{\alpha\beta},\nonumber \\
\end{eqnarray}
where $\epsilon^{\alpha}_{\bm{k^{||}}}=-2t_{\alpha}\left[\cos k_{x}+\cos k_{y}\right]$, $\bm{k^{||}}=(k_{x},k_{y},0)$ is defined as the transverse momentum, $\delta_{\alpha\beta}$ is the Kronecker delta function, and $\Sigma_{\alpha}(i \omega_{n})$ is the self energy on plane $\alpha$. Note that from this point on we will use the simplification that the hopping matrix elements are equal to $t$ for nearest neighbors, $t_{\alpha+1\alpha}=t_{\alpha-1\alpha}=t_{\alpha i j}=t$ and vanish otherwise. Since the EOM has a tridiagonal form with respect to the spatial component $z (\alpha,\beta)$ it can be solved with the so-called renormalized perturbation expansion\cite{econ}. We solve the equation directly, for the $\beta=\alpha$ case via
\begin{equation}
\label{green}
G_{\alpha\alpha}(i \omega_{n};\bm{k}^{||})=\frac{1}{i \omega_{n}+\mu_{\alpha}-V_{\alpha}+\Delta V_{\alpha}-\Sigma_{\alpha}(i \omega_{n})-\epsilon_{\bm{k}^{||}\alpha}+\frac{G_{\alpha-1\alpha}(i \omega_{n};\bm{k}^{||})}{G_{\alpha\alpha}(i \omega_{n};\bm{k}^{||})}t
+\frac{G_{\alpha\alpha+1}(i \omega_{n};\bm{k}^{||})}{G_{\alpha\alpha}(i \omega_{n};\bm{k}^{||})}t}.
\end{equation}
We create left and right recursion relations, 
\begin{equation}
\label{left}
L_{\alpha-n}(i \omega_{n};\bm{k}^{||})=i \omega_{n}+\mu_{\alpha}-V_{\alpha}+\Delta V_{\alpha}-\Sigma_{\alpha-n} (i \omega_{n})-\epsilon_{\bm{k}^{||}}+\frac{t^{2}}{L_{\alpha-n-1}(i \omega_{n};\bm{k}^{||})},
\end{equation}
and 
\begin{equation}
\label{right}
R_{\alpha+n}(i \omega_{n};\bm{k}^{||})=i \omega_{n}+\mu_{\alpha}-V_{\alpha}+\Delta V_{\alpha}-\Sigma_{\alpha+n}(i \omega_{n})- \epsilon_{\bm{k}^{||}}+\frac{t^{2}}{R_{\alpha+n+1}(i \omega_{n};\bm{k}^{||})}
\end{equation}
respectively to solve for the other values of $\alpha\neq\beta$. We start these relationships with the bulk values ($n\rightarrow \pm \infty$), which give us 
\begin{eqnarray}
L_{-\infty}(i \omega_{n};\bm{k}^{||})=\frac{i \omega_{n}+\mu_{\alpha}-V_{\alpha}+\Delta V_{\alpha}-\Sigma_{-\infty}(Z)-\epsilon_{\bm{k}^{||}}}{2}\nonumber \\
\pm \frac{1}{2} \sqrt{[i \omega_{n}+\mu_{\alpha}-V_{\alpha}+\Delta V_{\alpha}-\Sigma_{-\infty}(i \omega_{n})-\epsilon_{\bm{k}^{||}}]^{2}-4t^{2}}
\end{eqnarray}
and
\begin{eqnarray}
R_{\infty}(i \omega_{n};\bm{k}^{||})=\frac{i \omega_{n}+\mu_{\alpha}-V_{\alpha}+\Delta V_{\alpha}-\Sigma_{\infty}(i \omega_{n})-\epsilon_{\bm{k}^{||}}}{2}\nonumber \\
\pm \frac{1}{2} \sqrt{[i \omega_{n}+\mu_{\alpha}-V_{\alpha}+\Delta V_{\alpha}-\Sigma_{\infty}(i \omega_{n})-\epsilon_{\bm{k}^{||}}]^{2}-4t^{2}}.
\end{eqnarray}
The signs in the previous two equations are chosen to yield an imaginary part less than zero for $i \omega_{n}$ lying in the upper half plane, and vice versa for $i \omega_{n}$ lying in the lower half plane. The self-energies vanish for the ballistic metal leads used here. 

	To get our final expression for the Green's function, we substitute the left and right equations, Eqs. (\ref{left}) and (\ref{right}) respectively, into Eq. (\ref{green}) to yield
\begin{equation}
G_{\alpha\alpha}(i \omega_{n};\bm{k}^{||})=\frac{1}{i \omega_{n}+\mu-V_{\alpha}+\Delta V_{\alpha}-\Sigma_{\alpha}(i \omega_{n})-\epsilon_{\bm{k}^{||}\alpha}+L_{\alpha}(i \omega_{n};\bm{k}^{||})
+R_{\alpha}(i \omega_{n};\bm{k}^{||})}.
\end{equation}
	The local Green's functions on each plane can then be found by summing the Green's functions over the transverse momenta 
\begin{equation}
G_{\alpha \alpha}(i \omega_{n})=\int d\epsilon_{\bm{k}^{||}} \rho^{2D}(\epsilon_{\bm{k}^{||}}) G_{\alpha \alpha}(i \omega_{n},\epsilon_{\bm{k}^{||}})
\end{equation}
with
\begin{equation}
 \rho^{2D}(\epsilon_{\bm{k}^{||}})=\frac{1}{2 \pi^2 t a^2} K\left(1-\sqrt{1-\frac{(\epsilon_{\bm{k}^{||}})^{2}}{(4t)^{2}}}\right)
\end{equation}
being the $2D$ density of states (DOS), $K$ is the complete elliptical integral of the first kind and $a$ is the lattice constant which we set to $1$ for our calculations. 
After calculating the local Green's functions on each plane, we use Dyson's equation to define the effective medium for each plane,
\begin{equation}
G^{-1}_{0 \alpha}(i \omega_{n})=G^{-1}_{\alpha}(i \omega_{n})+\Sigma_{\alpha}(i \omega_{n}).
\end{equation}

The local Green's function for the $\alpha$-th plane then satisfies
\begin{equation}
G_{\alpha}(i \omega_{n})=\frac{(1-w_{1})}{G^{-1}_{0 \alpha}(i \omega_{n})+\frac{1}{2}U_{\alpha}}+ \frac{w_{1}}{G^{-1}_{0 \alpha}(i \omega_{n})^{-1}-\frac{1}{2}U_{\alpha}}
\end{equation}
where $w_{1}$ is the average filling of the localized particles.  Finally, we use the new local Green's functions and Dyson's equation again to find the self-energy,
\begin{equation}
\Sigma_{\alpha}(i \omega_{n})=G^{-1}_{0 \alpha}(i \omega_{n})-G_{\alpha}^{-1}(i \omega_{n}).
\end{equation}
This forms the basic algorithm for dynamical mean-field theory, which we now augment to determine the capacitance. 

To calculate the capacitance, we need the quantum-mechanically calculated electron number density and the potential at plane $\alpha$ . We calculate the electronic charge on each plane by summing the Green's functions over all Matsubara frequencies on the imaginary axis, multiplied by the temperature. The electron number density at plane $\alpha$ satisfies
\begin{equation}
\rho_{\alpha}=\frac{1}{2}+T \sum_{n}G_{\alpha}(i \omega_{n}).
\end{equation}
We can take advantage of the behavior of $G_{\alpha}(i \omega_{n})$ at large $n$, which goes like $1/i \omega_{n}$, allowing us to regularize the Matsubara frequency summation by adding and subtracting $T \sum_{n}{1/[i \omega_{n}+\mu- Re\Sigma_{\alpha}(i \omega_{n_{max}})]}$. This gives an exact summation of the tail of the Matsubara sums and the electron number density becomes
\begin{eqnarray}
\rho_{\alpha}=\frac{1}{2}+T \sum_{n}\left[G_{\alpha}(i \omega_{n})-\frac{1}{i \omega_{n}+\mu_{\alpha}- \rm{Re}\Sigma_{\alpha}(i \omega_{n_{max}})} \right]\nonumber\\
-\frac{1}{2} \tanh \left[\frac{\beta[\mu_{\alpha}- \rm{Re}\Sigma_{\alpha}(i \omega_{n_{max}})]} {2}\right].
\label{eq:rhoalpha}
\end{eqnarray}

To find the Coulomb potential on each plane we begin with the magnitude of the local electric field created on plane $\alpha$,
\begin{equation}
\left|\textbf{E}\right| = \frac{\left|e\right| \left|\rho_{\alpha}-\rho^{bulk}_{\alpha}\right|a}{2\epsilon_{0} \epsilon_{r \alpha}}.
\end{equation}
where $e$ is the charge of an electron, $\epsilon_{0}$ is the permittivity of free space, $\epsilon_{r \alpha}$ is the relative permittivity of plane $\alpha$, and $\rho^{bulk}_{\alpha}$ is the bulk electron density of the material of which plane $\alpha$ is composed. Once the total field is known for each plane, we integrate them to find the electric potentials. 
	Since the electric field's magnitude is constant, it is straightforward to compute the Coulomb potential, 
\begin{equation}
V_{\beta}(\alpha) = -\sum_{\alpha} (\rho_{\alpha}-\rho^{bulk}_{\alpha}-\bar{\rho}) \left\{ \begin{array}{ll}
  \sum^{\beta}_{\gamma=\alpha+1}\frac{1}{2}\left[e_{Schot}(\gamma)+e_{Schot}(\gamma-1)\right], &\mbox{$\beta>\alpha$} \\
  0, &\mbox{$\beta=\alpha$} \\
  \sum^{\gamma=\alpha-1}_{\beta}\frac{1}{2}\left[e_{Schot}(\gamma)+e_{Schot}(\gamma+1)\right], &\mbox{$\beta<\alpha$}
       \end{array} \right.
       \label{eq:V}
\end{equation}
where we define the parameter,
\begin{equation}
e_{Schot}(\alpha )= \frac{e^{2}a}{2\epsilon_{0} \epsilon_{r \alpha}},
\end{equation}
which characterizes the decay of the surplus charge density away from the interface. The parameter $\bar{\rho}=\sum_{\alpha}{(\rho_{\alpha}-\rho^{bulk}_{\alpha})}/N$ (with $N$ the total number of self-consistent planes used in our calculations) is used to improve the convergence of our equations and vanishes for the converged final fixed-point solution. It is worth noting that we are fixing $e_{Schot}$ (equal in both the metal leads and barrier) and not recalculating the dielectric constant, meaning the many-body effects on the dielectric are already incorporated \textit{ a priori} in the calculation. The input $e_{Schot}$ incorporates all contributions to the dielectric, including the bare dielectric, ion core, etc., therefore we can not directly compare our results to the geometric capacitance in Eq. (\ref{eq:geom}) because we cannot isolate the different contributions to the dielectric to find the effective $\epsilon_{r}$ needed in the formula for the geometric capacitance. 

	We can now state the full algorithm used in our calculations. We start by inputting a value for the screening length ($e_{Schot})$, the applied potential ($\Delta V=V_{a}/2$), the chemical potential ($\mu$), and the temperature ($T$). We begin the iterative calculations with a guess for the self-energy on each plane, usually zero or the solution to a previous calculation. Next, we use the left and right recursive equations to calculate the local Green's functions at each plane. These local Green's functions are then used to calculate the effective medium for each plane, which in turn is used to solve for the impurity Green's functions. The new impurity Green's functions are used to calculate the new self-energies which are used to feed the loop again. Additionally the impurity Green's functions are used to extract the planar filling. The planar filling is used within classical electrostatics to calculate the electric potential on each plane and in turn the contribution of the potential energy to the electrochemical potential on each plane. We average the new potentials,  
	\begin{equation}
V^{next\; iteration}_{\alpha}=\alpha_{V} V^{old}_{\alpha}+(1-\alpha_{V})V^{new}_{\alpha}
\end{equation}
with a large damping factor $\alpha_{V}$. The parameter $\alpha_{V}$ is usually at least $0.99$ which is needed to slow the updating and prevent converging to a nonphysical solution. We iterate through these steps until the calculations converge. Due to the large damping factor, it typically takes between 1,000 and 10,000 iterations to reach convergence. In order to achieve proper convergence the errors are kept to less than one part in $10^5$, this allows for reproducibility in the algorithm ensuring the planar charge densities are accurately calculated from one iteration to the next. Keeping the error below a tolerable level has proven difficult to achieve in models other than the Falicov-Kimball model (such as the Hubbard model).

After the calculations reached a self consistent solution we can use our two methods to extract the capacitance. The CoC capacitance is calculated by summing over the extracted planar filling [Eq.(\ref{eq:geom2}) for both the right and left halves of the layers]. The VP capacitance is calculated by substituting the calculated electric potential at the left and right interface layers in for $V_{L}$ and $V_{R}$, respectively [Eq. (\ref{eq:modcap})].

\begin{center}
\textbf{III. RESULTS}
\end{center}

	All of our numerical results will be calculated at half-filling ($\mu=0,\langle c^{\dagger}_{i}c_{i} \rangle=1/2$, and $w_{1}=\langle w_{i} \rangle =1/2$). We build our model with 30 self-consistent metal planes in the leads each terminating in the bulk surrounding the dielectric layers in the center. We vary the thickness of the dielectric region from 4 to 20 planes. The calculations are carried out on a simple cubic lattice allowing only nearest neighbor hopping (both interplane and intraplane hopping, $t$, are equal). This reduces the number of parameters in the calculations allowing focus on the properties of interest.

\begin{figure}[htp]
	\centering
		\includegraphics[width=120mm]{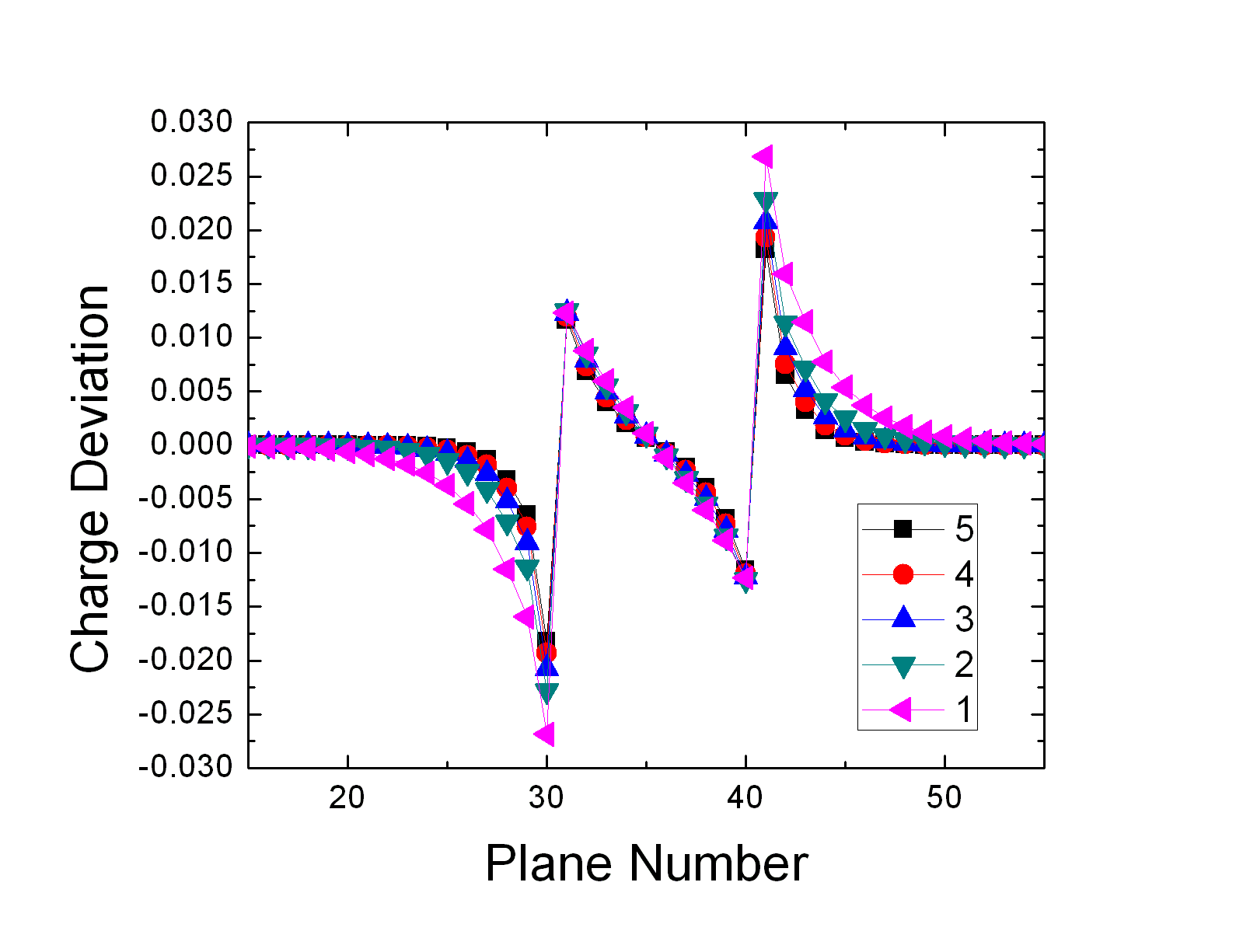}
		\includegraphics[width=120mm]{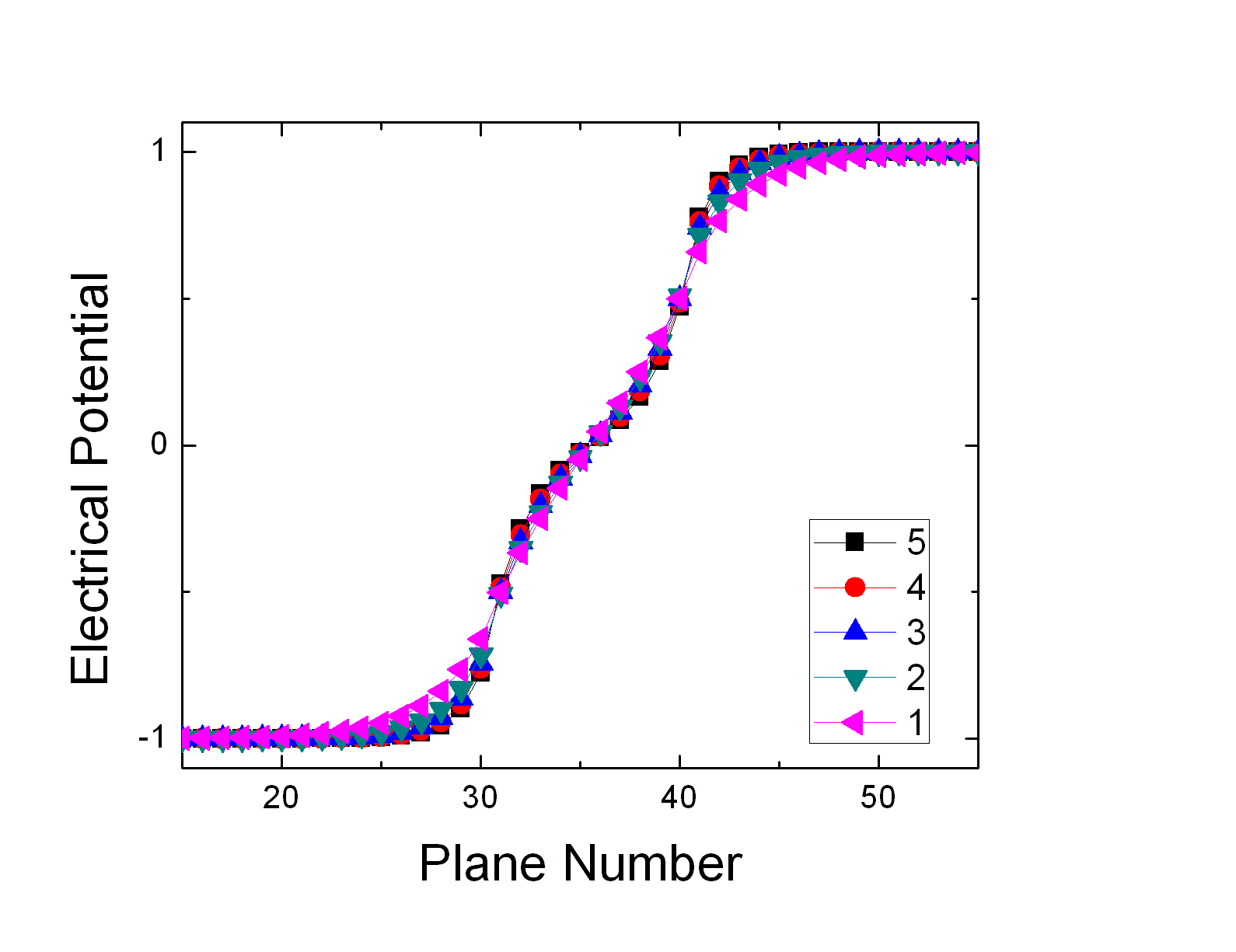}
	\caption{(Color online) [10 dielectric planes, $U=6$, $w_{1}=0.5$, $E_{f}=0$, $V_{a}=\left|2\right|$, and $T=0.25$] Charge deviations from the bulk charge (a) and the electric potential (b) plotted as a function of plane numbers for various $e_{Schot}$ as indicated by the legends.}
\label{fig:Eschotcharge}
\end{figure}

	There are many parameters that can be varied to investigate their effects on the capacitance. For each set of parameters in our calculations, we can extract the quantum-mechanically calculated electron number density for each plane from Eq. (\ref{eq:rhoalpha}), as seen in Fig. \ref{fig:Eschotcharge} (a), which plots the difference between the electron number density and the bulk electron number density through the device for various $e_{Schot}$. From Eq. (\ref{eq:V}) we can also plot the potentials on each plane through the device, which is shown in Fig. \ref{fig:Eschotcharge} (b), again for various values of $e_{Schot}$. Note how the most rapid change in the potentials occur near the interface. With the charge and the potentials known, we can calculate the capacitance for each device. Once the potentials have been calculated through the devices the capacitance is calculated from Eq. (\ref{eq:modcap}). Figure \ref{fig:Eschotcharge} (a) and Figure \ref{fig:Eschotcharge} (b) show that as the screening length increases the charge deviation curves become, as expected, sharper in nature. 
	
\begin{figure}[htp]
	\centering
		\includegraphics[width=120mm]{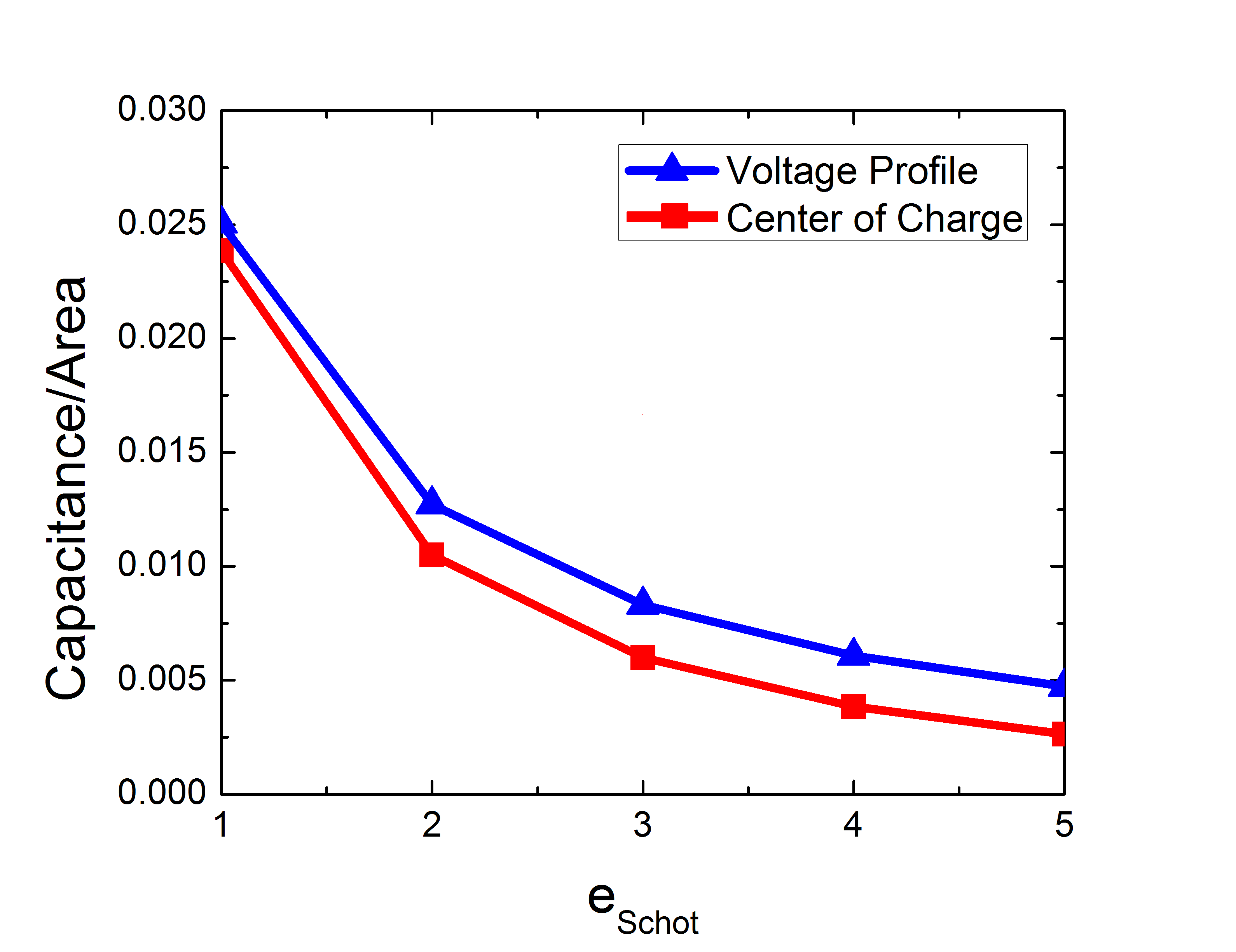}
	\caption{(Color online) [10 dielectric planes, $U=6$, $w_{1}=0.5$, $E_{f}=0$, $V_{a}=\left|2\right|$, and $T=0.25$] The capacitance per unit area plotted as a function of $e_{Schot}$.}
\label{fig:Eschotcap}
\end{figure}

Once we have calculated the capacitance per unit area we can compare the center of charge capacitance as defined in Eq. (\ref{eq:geom2}) with the voltage profile capacitance calculated from Eq. (\ref{eq:modcap}). Both methods of calculating the capacitance are plotted in Fig. \ref{fig:Eschotcap} which shows that as we vary the screening length, both formulas for the capacitance follow the expected $C/A \propto (1/e_{Schot})$ behavior.

\begin{figure}[htp]
	\centering
		\includegraphics[width=120mm]{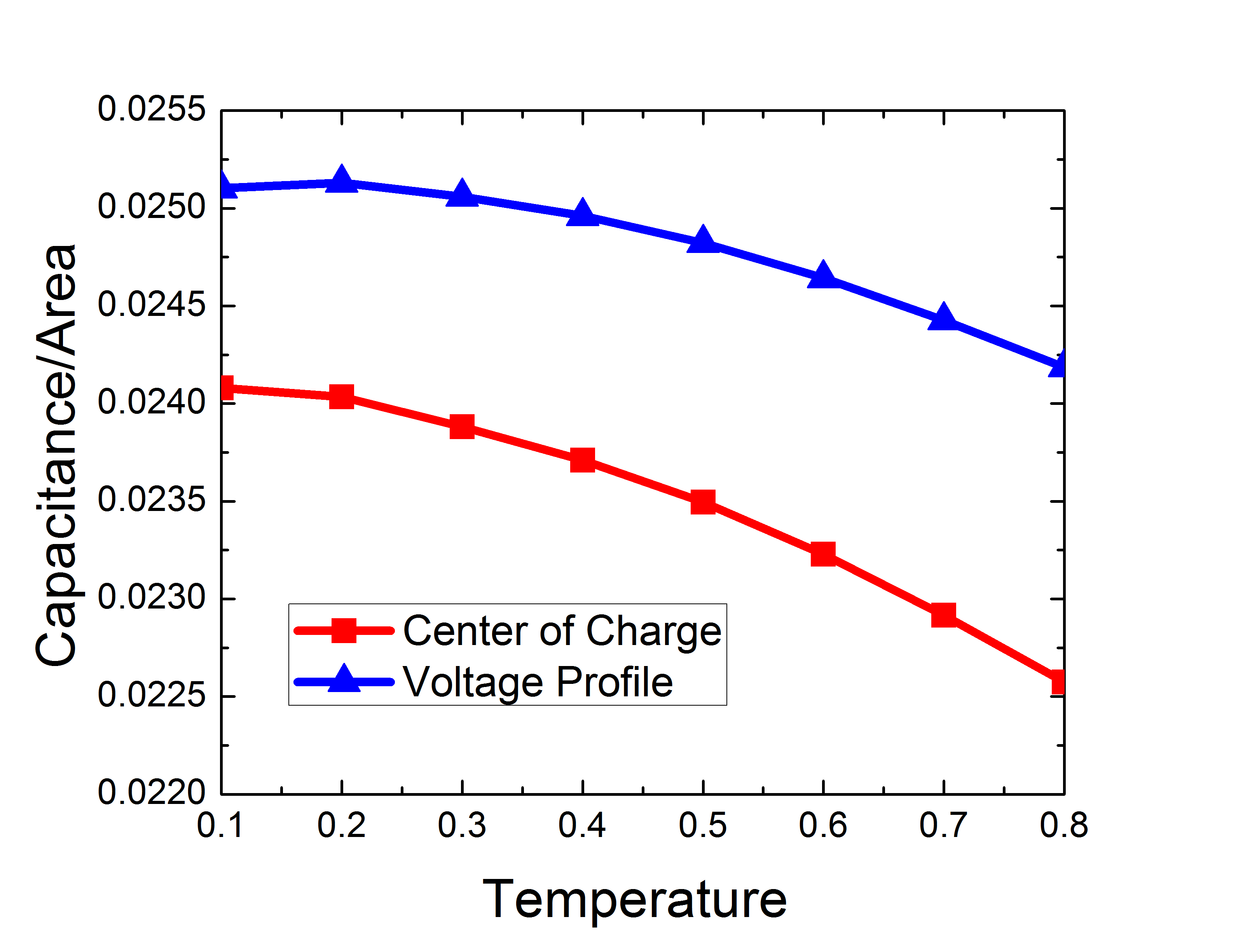}
	\caption{(Color online) [10 dielectric planes, $e_{Schot}=1$, $U=6$, $w_{1}=0.5$, $E_{f}=0$, and $V_{a}=\left|1\right|$] The capacitance per unit area plotted as a function of temperature.}
\label{fig:Temperature}
\end{figure}

 	We investigate thermal effects on the capacitance by varying the temperature. Figure \ref{fig:Temperature} shows that the capacitance/area falls off slightly ($\approx4\%$) as the temperature is increased over the range of $0.1$ to $0.8$.  In our calculation, if we take a reasonable energy scale for our system, such as a noninteracting bandwidth of 3 eV, then $t=0.25$eV ($U=6$). This corresponds to a temperature range from room temperature ($T=0.1$) to 2500K ($T=0.8$). 
 	
\begin{figure}[htp]
	\centering
		\includegraphics[width=120mm]{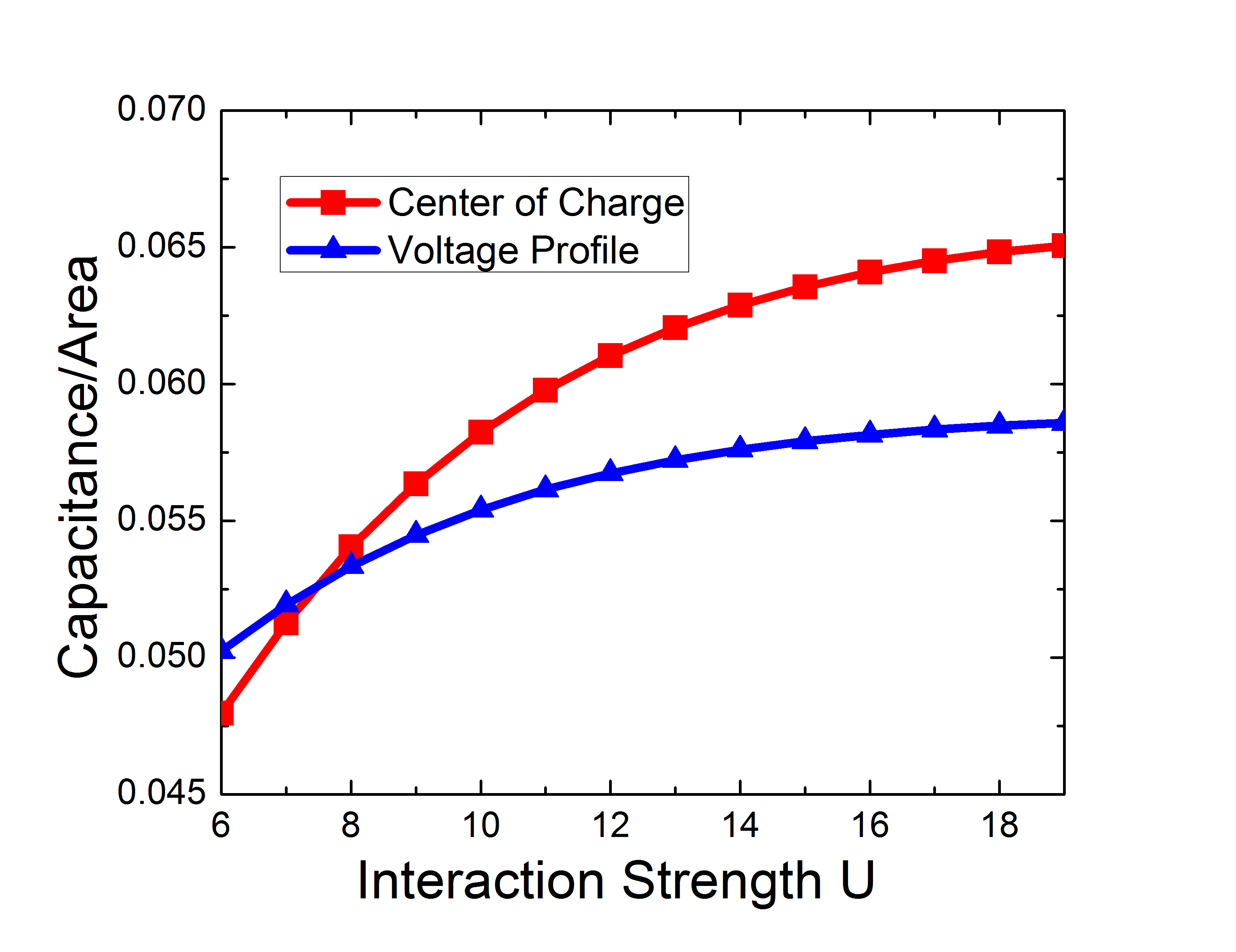}
	\caption{(Color online) [10 dielectric planes, $e_{Schot}=0.5$, $w_{1}=0.5$, $E_{f}=0$, $V_{a}=\left|1\right|$, and $T=0.25$] Capacitance per unit area for various interaction strengths $U$, with a crossover seen at approximately $U=7.5$}
\label{fig:UCap}
\end{figure}

	Fig. \ref{fig:UCap} shows that the capacitance per unit area has a stronger dependence on the Falicov-Kimball interaction strength ($U$). The CoC capacitance per unit area grows faster than the VP capacitance per unit area. The CoC capacitance per unit area increases by approximately $30\%$ across the range of interaction strengths.  There is a crossover around $U=7.5$ where the center of CoC becomes larger than the VP capacitance. The increase in capacitance that results from increasing the interaction strength in the barrier is due to the higher interaction strength reducing the spatial extent of the dipole layer that is formed on the inside of the dielectric region.
	
\begin{figure}[htp]
	\centering
		\includegraphics[width=120mm]{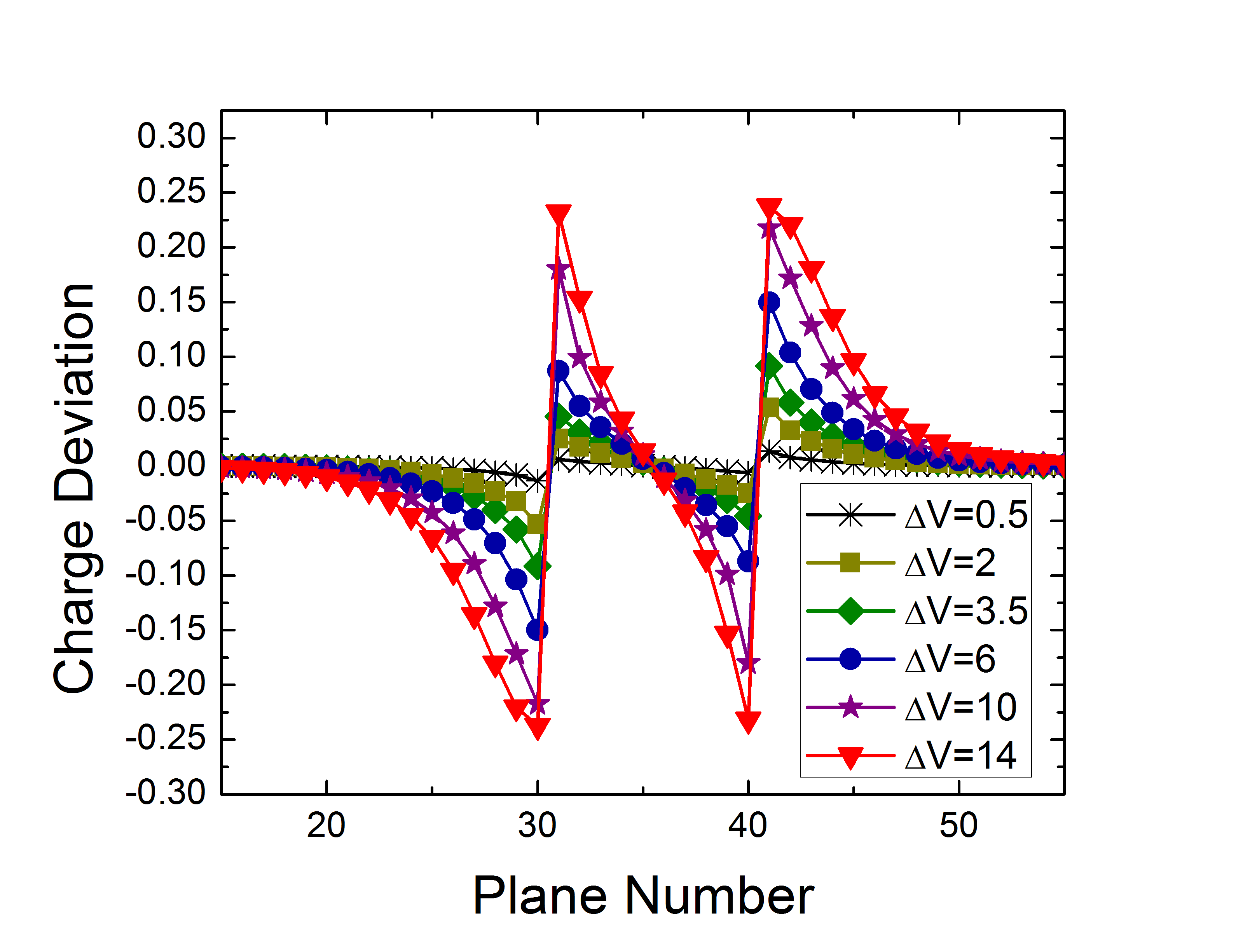}
		\includegraphics[width=120mm]{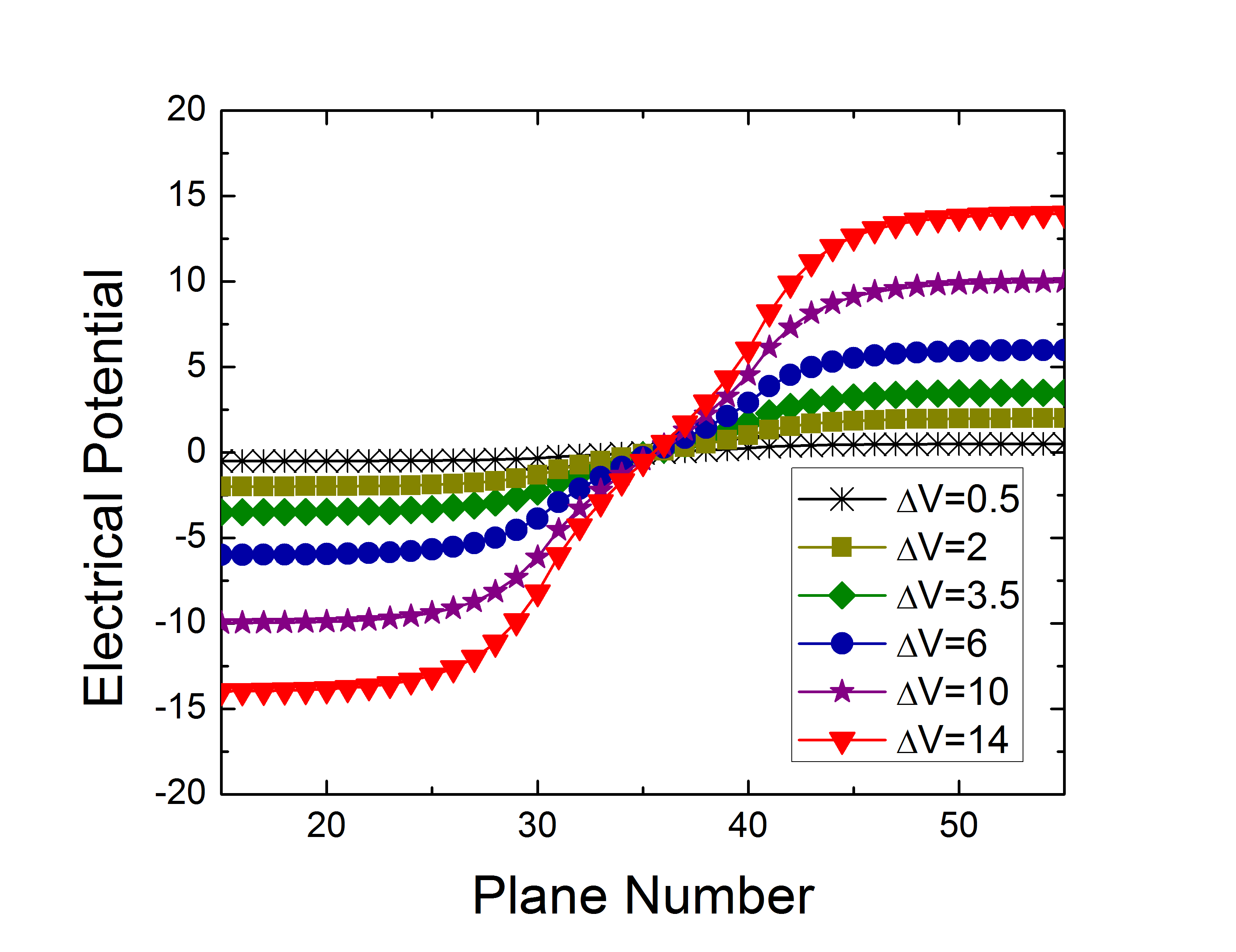}
	\caption{(Color online) [10 dielectric planes, $e_{Schot}=1$, $U=6$, $w_{1}=0.5$, $E_{f}=0$, and $T=0.25$]  Charge deviation from the bulk (a) and the electrical potential (b) profile for various applied potentials.}
\label{fig:VCD}
\end{figure}

\begin{figure}[htp]
	\centering
		\includegraphics[width=120mm]{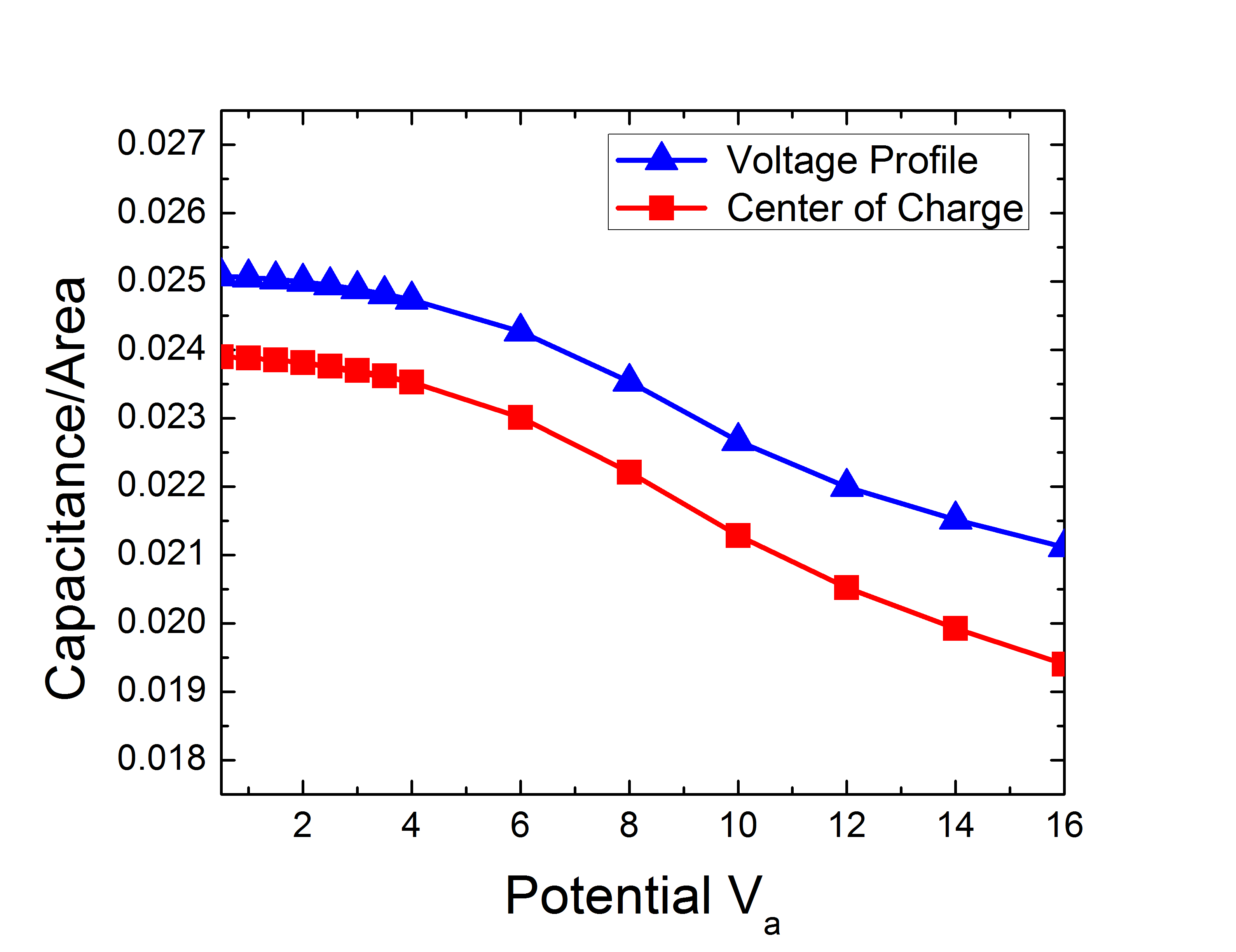}
	\caption{(Color online) [10 dielectric planes, $e_{Schot}=1$, $U=6$, $w_{1}=0.5$, $E_{f}=0$,  and $T=0.25$] Capacitance/Area for various applied potentials, $V_{\alpha}$.}
\label{fig:VCap}
\end{figure}

	We plot the charge deviation and potential profiles for various applied potentials ($V$) in Fig. \ref{fig:VCD} as well as the capacitance/area for various applied potentials in Fig. \ref{fig:VCap}. The reduction in capacitance due to increasing the applied potential can be seen in  Fig. \ref{fig:VCD} as the planes on either side of the dielectric layer begin to saturate and excess charge is forced further away from the interface leading to a reduction in the capacitance. 

\begin{center}
\textbf{IV. CONCLUSIONS}
\end{center}

	We presented in this paper a self-consistent method for using IDMFT to calculate the capacitance of multilayered nanostructures. We also discussed the various capabilities and challenges with this many-body approach. We showed how using the quantum zipper algorithm based on the work of Potthoff and Nolting, we can calculate the electron number densities on each plane. The electron number densities were then used to calculate the electric potential on each plane from classical electrostatics. We presented two methods to calculate the capacitance, one based on a center of charge approach and the other accounting for the screening of the charges. The capacitance is calculated for various parameters, finding the strongest dependence on the interaction strength (U), 30\% over the calculated range. We find a weaker dependence on temperature (5-10\%) and applied potential.  

	We use a semiclassical approach to calculate the potential, which fixes the relation between the applied potential and the electronic charge density at a given plane. By making the dielectric a fixed parameter we cannot capture any effects on the capacitance that are not already described in Eq. (\ref{eq:geom2}) or Eq. (\ref{eq:modcap}).  Fixing the dielectric in the system forces the parameter to describe the total dielectric not just the dielectric values coming from the polarizability of the ion cores. Although this limits our ability to isolate many-body effects in Eq. (\ref{eq:geom2}) or Eq. (\ref{eq:modcap}) from the geometric capacitance in Eq. (\ref{eq:geom}), we are still able to examine the behavior of the model as we vary other parameters. 

	We calculated the capacitance via two methods, through the generated voltage profiles, $C_{VP}$, as well as the center of charge approach, $C_{CoC}$. Calculating the capacitance in these methods allows for comparison to different experimental setups. For example, if the capacitance of an experimental set up is determined by integrating the total charge required to discharge the capacitor, then comparison to the $C_{CoC}$ method would be more appropriate because the total charge is $\sum_{\alpha}\rho_{\alpha}$ and we are not measuring the potential at the capacitor plates but between the left and right leads. On the other hand, if the experimental set up had the ability to probe the potential at the edge of the metallic plates, then the $C_{VP}$ method is more appropriate for comparison, since we do not measure the total charge.

	The ability to calculate the potential and charge profiles that incorporate many-body effects allow us to investigate any non-linear effects. The calculations for the capacitance presented in this paper were carried out in the slow limit with no current flow {\cite{Lutt}}, reproducing the inverse dependence on thickness and linear dependence on $e_{Schot}$.  In future research moving away from half-filling would allow for the model to hopefully capture more non-linear behavior as the Falicov-Kimball model enters into a phase-separated state. Other potentially interesting and non-linear behavior can be investigated by using the Hubbard model rather than the Falicov-Kimball model, which is currently inhibited by the achievable accuracy of the required numerical calculations. 

	The experiments see the greatest enhancements to the capacitance near the highly depleted limit, therefore by moving the Falicov-Kimball model away from half filling and forcing the model to enter a phase-separated state, we expect to see enhancements in our model. The many body effects should become more pronounced as a phase-separated state is entered.

\begin{center}
\textbf{ACKNOWLEDGMENTS}
\end{center}
We acknowledge the support of the National Science Foundation through Grant No. DMR-1006605. J. K. Freericks acknowledges support by the McDevitt bequest at Georgetown. We also acknowledge useful conversations with Thilo Kopp and Jochen Mannhart.

\end{document}